\pdfoutput=1
\documentclass[twocolumn]{aastex6} %

\usepackage{natbib}
\usepackage{amsmath}
\usepackage{amssymb}
\usepackage{subfigure}
\usepackage{url}
\usepackage{CJK}

\renewcommand{\vec}[1]{ {\mathbf #1} }

\newcommand{\curl}{ {\bf \nabla} \times}

\newcommand{\Fig}{{Figure}}

\shorttitle{Reconstruction of Coronal Current Sheet}
\shortauthors{Jiang et al.}

\begin{document}
\begin{CJK*}{UTF8}{gbsn}

  \title{Reconstruction of a Large-scale Pre-flare Coronal Current
    Sheet Associated with an Homologous X-shaped Flare}

\author{
  Chaowei Jiang\altaffilmark{1,2,4},
  Xiaoli Yan\altaffilmark{3},
  Xueshang Feng\altaffilmark{4,1},
  Aiying Duan\altaffilmark{5},
  Qiang Hu\altaffilmark{2},
  Pingbing Zuo\altaffilmark{1,4},
  Yi Wang\altaffilmark{1,4}}

\altaffiltext{1}{Institute of Space Science and Applied Technology,
  Harbin Institute of Technology, Shenzhen 518055, China, chaowei@hit.edu.cn}

\altaffiltext{2}{Center for Space Plasma and Aeronomic Research, The
  University of Alabama in Huntsville, Huntsville, AL 35899, USA}

\altaffiltext{3}{Yunnan Observatories, Chinese Academy of Sciences,
  396 Yangfangwang, Guandu Districk, Kunming 650216, Yunnan,
  P. R. China}

\altaffiltext{4}{SIGMA Weather Group, State Key Laboratory for Space
  Weather, National Space Science Center, Chinese
  Academy of Sciences, Beijing 100190, China}

\altaffiltext{5}{Key Laboratory of Computational Geodynamics, University
  of Chinese Academy of Sciences, Beijing 100049, China}

\begin{abstract}
  As a fundamental magnetic structure in the solar corona,
  electric current sheets (CSs) can form either prior to or during
  solar flare, and they are essential for magnetic energy dissipation
  in the solar corona by enabling magnetic reconnection. {However static reconstruction of
  CS is rare, possibly due to limitation inherent in available coronal field extrapolation codes}. Here we present the
  reconstruction of a large-scale pre-flare CS in solar active region~11967
  using an MHD-relaxation model constrained by
  SDO/HMI vector magnetogram. The CS is found to be associated with a set of peculiar homologous
  flares that exhibit unique X-shaped ribbons and loops occurring in a
  quadrupolar magnetic configuration. {This is evidenced by that} the field lines traced from the CS to
  the photosphere form an X shape which nearly precisely reproduces the
  shape of the observed flare ribbons, {suggesting} that the flare
  is a product of the dissipation of the CS through
  reconnection. The CS forms in a hyperbolic flux tube, which is an intersection of two
  quasi-separatrix layers. The recurrence of the X-shaped flares {might} be attributed
  to the repetitive formation and dissipation of the CS, as driven by the photospheric footpoint motions.
  These results demonstrate the power of data-constrained MHD model
  in reproducing CS in the corona as well as providing insight into
  the magnetic mechanism of solar flares.
\end{abstract}


\section{Introduction}
\label{sec:intro}

The Sun has a complex dynamo that generates electric currents and
magnetic fields, which provide energy to heat the corona and power solar flares and
eruptions. The currents in solar corona are typically evolved into
two forms, volumetric channels (manifested as twisted magnetic flux
ropes) and narrow sheets across which the magnetic field vector is
almost discontinuous. Current sheets (CSs) are essential for energy
dissipation in the solar corona, in particular by enabling magnetic
reconnection. For instance, the large-scale vertical CS is a basic
building block in the standard flare model~\citep{LinJ2015}, which
extends from the top of post-flare loops to the bottom of an erupting
flux rope, and in which reconnection continuously occurs.
CS does not only exist during flare/eruption but also form in
the quasi-static evolution of the corona subjected to the slow
photospheric motions. The Parker theory demonstrates that the
evolution of coronal magnetic fields in response to slow photospheric
footpoint motions in general produces states with CSs rather than
smooth force-free equilibria, that is, such a spontaneous formation of
electric current sheets is a basic magnetohydrodynamics (MHD)
process~\citep{Parker1972, Parker1994, LowBC1996, Low2005,
  Low2010}. {This means that the presence of CSs should be ubiquitous in the corona, like
  the presence of flux ropes.} CS can form in either magnetic separartix surfaces, which
define the boundary of topologically separated domains (i.e., the
magnetic field-line mapping is discontinuous), or quasi-separatrix
layers (QSLs), where magnetic field-line mapping has a steep yet
finite gradient~\citep{Demoulin2006, Aulanier2006}.

However, in reconstructing realistically the coronal magnetic field using existing models, even
including the up-to-date most sophisticated nonlinear force-free field (NLFFF)
models~\citep[e.g.,][]{DeRosa2009, Wiegelmann2012solar},
the reconstruction of CS is rarely reported in the literature.
On the other hand, it is well known that many NLFFF models are able to reconstruct
magnetic flux ropes in the corona~\citep{GuoY2017, ChengX2017}.
Considering the ubiquitous presence of CS like flux rope, there might be some problems~\citep[for example, see][]{Low2013} with available NLFFF reconstruction codes, which make them fail to reproduce CS.
Regarding that formation of CS is very natural in MHD relaxation process,
it might be more suitable to use an MHD-relaxation method to reconstruct
the coronal field containing CSs than many NLFFF codes that are based on mainly non-MHD approaches.
From a theoretical point of view, the CS is a discontinuity which constitutes weak solution of the MHD equation, and thus numerical reconstruction of CS is a task of obtaining such weak solution. Since the MHD-relaxation method is based on well-developed computational-fluid-dynamics (CFD) theory and codes, weak solutions are allowed by the CFD codes and can be reproduced correctly. However, the weak solution theory (i.e., solution contains discontinuities) is not yet established for typical NLFFF codes like the optimization and Grad-Rubin ones. So, it is problematic whether those NLFFF codes can produce solutions with CS discontinuities.


In this paper, we show that CSs in the corona could indeed be reconstructed
using an MHD relaxation model constrained by vector
magnetograms. We investigated a homologous X-shaped flare in AR~11967
and found that the MHD equilibrium prior to the flares includes a
large-scale CS situated vertically in the center of the flare site,
which has a magnetic quadrupolar configuration. Furthermore, the
photospheric footpoints of the field lines traced from the CS match
strikingly well with the X-shaped flare ribbons, which provides strong
evidence that reconnection in the CS produces the flares.


\begin{figure}[htbp]
  \centering
  \includegraphics[width=0.48\textwidth]{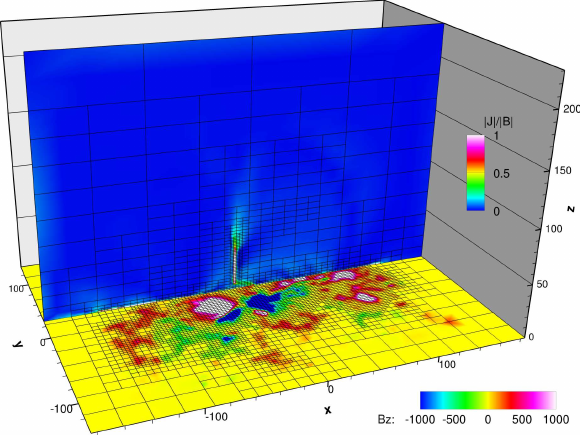}
  \caption{Structure of the AMR grid. Six levels of refinement are used
    in this computation, and in each level the grid size is refined by
    a factor of 2. The lower slice of $z=0$ is shown with the
    distribution of $B_{z}$ on the photosphere and the vertical slice
    (pseudo-colored by $J/B$) is a cross section of the 3D volume. The
    grid lines show the structure of the grid blocks and each block
    consists of $8^{3}$ cells which is omitted here. Unit of length is
    Mm. The highest resolution in $x-y$ plane is 0.36~Mm.
    Note that the resolution in vertical direction is higher by a
    factor of 2 than that in the horizontal direction. This is
    designed considering that the magnetic field often expands more
    strongly in the vertical direction than in horizontal direction.}
  \label{fig:amr_mesh}
\end{figure}
\begin{figure*}[htbp]
  \centering
  \includegraphics[width=0.95\textwidth]{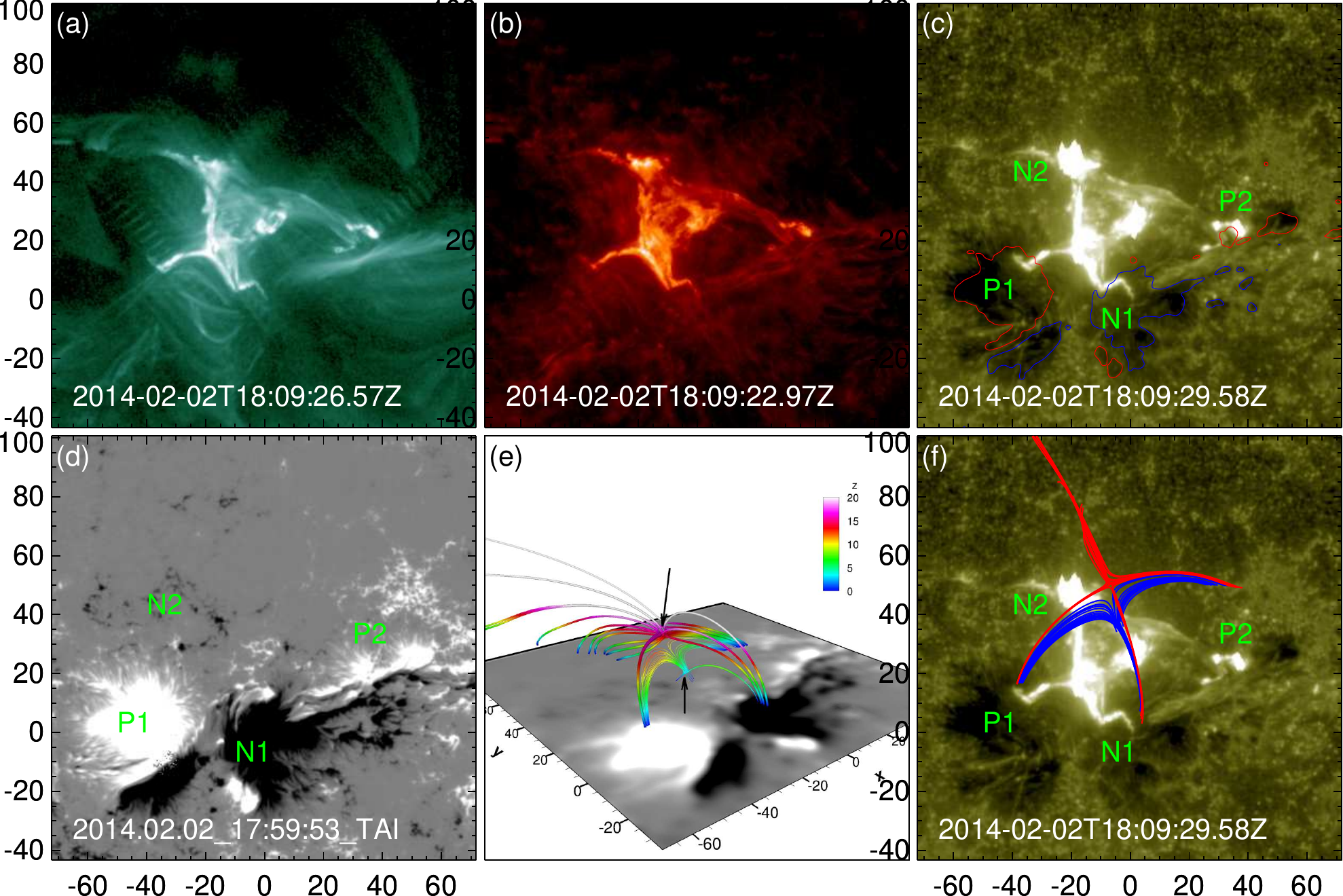}
  \caption{SDO observations of the X-shaped M3.1 flare at
    18:11~UT on 2014 February 2. (a), (b) and (c) are snapshots of the
    flare at main phase in AIA 94~{\AA}, 304~{\AA}, and 1600~{\AA}. (d)
    The photospheric magnetic flux distribution (magnetogram) observed
    by HMI at 10~min prior to the flare peak time. Four polarities
    associated with the flare are labeled as P1, N1, P2, and N2. Here
    P1 and N1 denote the major sunspots defining AR~11967; P2 refers
    to the positive flux region in the north of N1, including the
    plage region as well as the elongated positive polarity of strong
    field, while N2 denotes the plage region in the north of P1.
    Overlaid contours in (c) are photospheric magnetic field of
    $1000$~G (red) and $-1000$~G (blue). (e) Magnetic field lines of a
    potential field extrapolation from the magnetogram. The two arrows
    mark the locations of two magnetic null points of the potential
    field. These two nulls are close to each other in the horizontal
    direction, one situating at altitude of $\sim$4~Mm and the other
    at $\sim$18~Mm. The field lines are traced in the neighborhood of
    the nulls, with the thick ones for the null point higher while the
    thin ones for the null lower. (f) The double-null related field
    lines overlaid on the AIA 1600~{\AA} image. Lines of the higher
    (lower) null is colored as red (blue). The unit of length in all
    the panels is Mm.}
  \label{fig:4aia_hmi}
\end{figure*}

\begin{figure*}[htbp]
  \centering
  \includegraphics[width=0.8\textwidth]{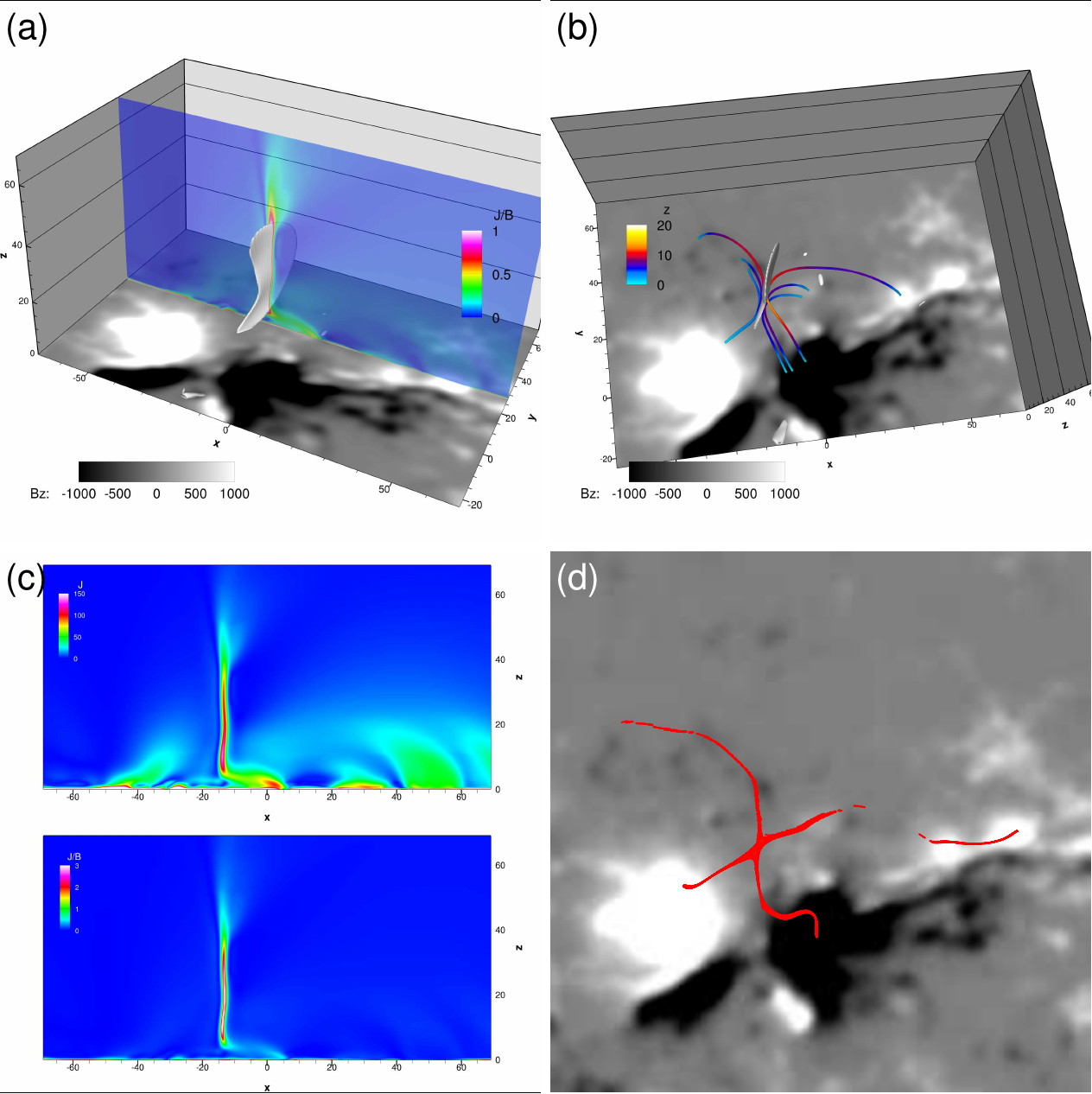}
  \caption{Structure of the CS related with the X-shaped flare at
    18:11~UT. (a) The 3D gray object is iso-surface of
    $J/B=1$~Mm$^{-1}$ which is used to define the CS. A vertical cross
    section of the volume is shown with distribution of $J/B$. (b)
    Another view of the CS. The sampled field lines are traced from
    the two sides of the CS, and the color represents the height
    $z$. (c) Distribution of current density $J$ and $J/B$ on the
    vertical cross section shown in (a). Here $J=|\curl \vec B|$ and
    its unit is G~Mm$^{-1}$. (d) The red dots are footpoints of all
    the closed field lines that are traced from the CS, and they are overlaid on the
    magnetogram. A total number of 2469 field lines are traced. The
    unit of length in all the panels is Mm. }
  \label{fig:CS_structure}
\end{figure*}
\begin{figure*}[htbp]
  \centering
  \includegraphics[width=0.8\textwidth]{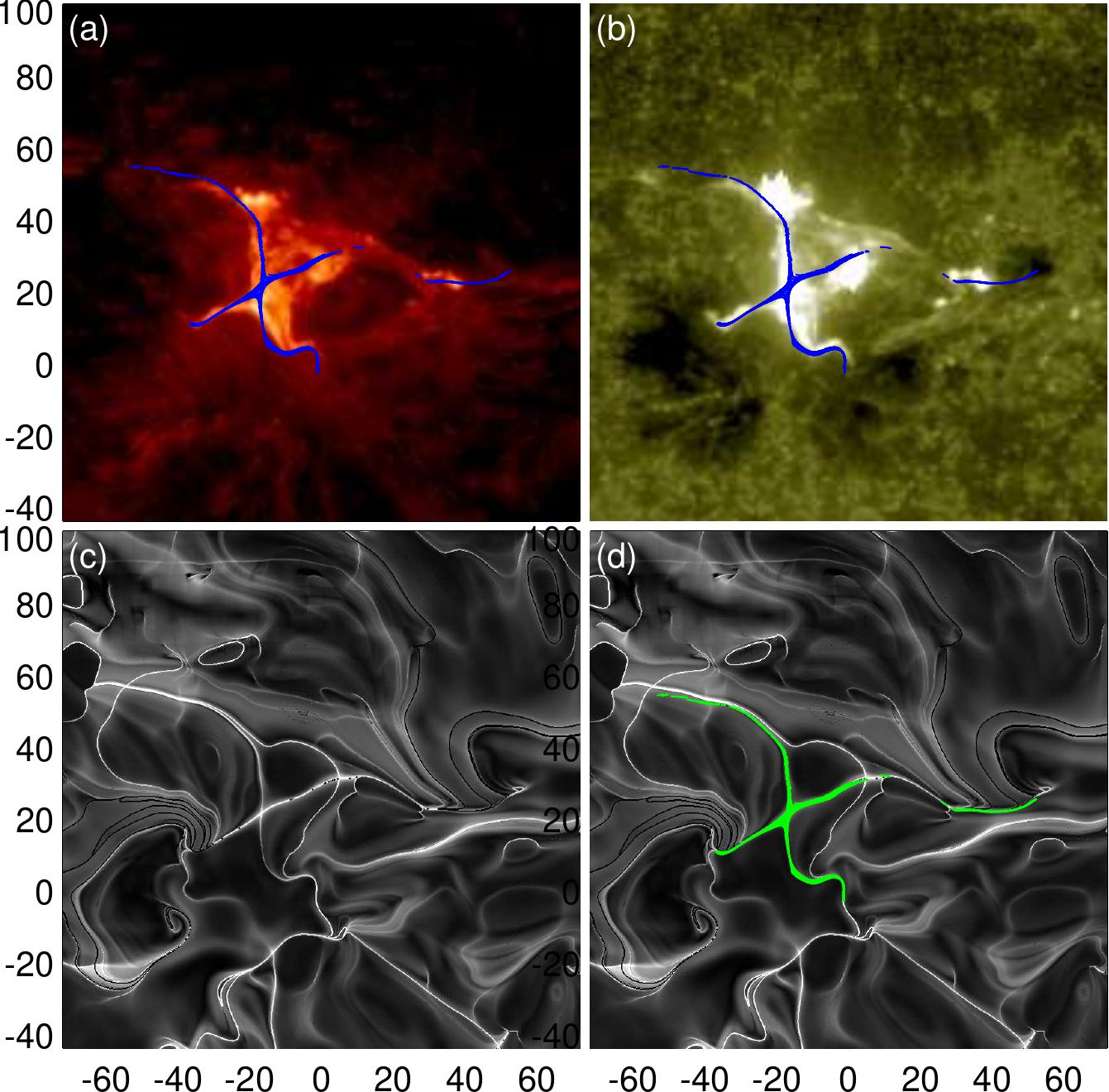}
  \caption{Comparison of the modeled results with the observed flare
    ribbons. Panels (a) and (b) are respectively AIA 304~{\AA} and 1600~{\AA}
    images overlaid with the footpoints of magnetic field lines
    contacting the CS (in blue dots). (c) Map of magnetic squashing
    degree $\log Q$.  The color range from black to white represents 0
    to 5. (d) The same footpoints in (a) and (b) overlaid on the
    squashing-degree map.}
  \label{fig:ribbon_compare_qsl}
\end{figure*}

\begin{figure*}[htbp]
  \centering
  \includegraphics[width=0.98\textwidth]{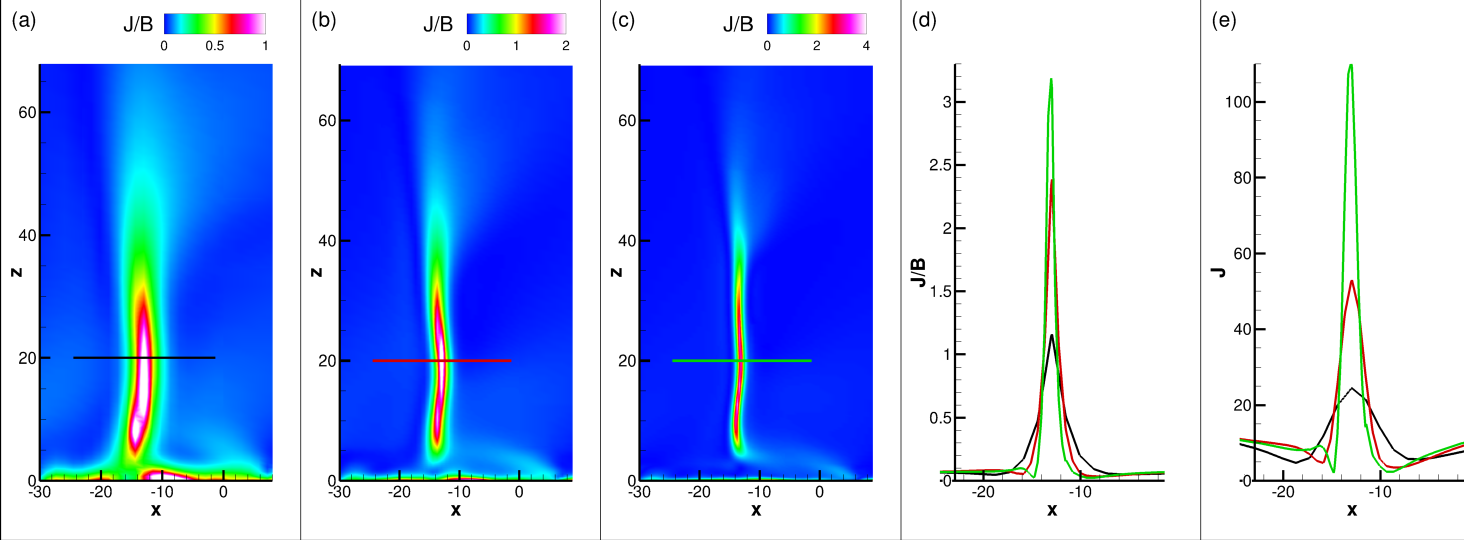}
  \caption{Reconstructing the CS using three different grid
    resolutions with finest grid in horizontal direction of 1.44,
    0.72, 0.36~Mm, respectively. (a), (b), and (c) are the vertical
    cross sections of the CS. The cross sections have the same
    location shown in \Fig~\ref{fig:CS_structure}a and c. (d) and (e)
    shows a 1D profile of $J/B$ and $J$ on the same horizontal line
    ($z=20$) for the three resolutions. The units for $J/B$ and $J$
    are respectively, Mm$^{-1}$ and G~Mm$^{-1}$, and the length unit
    is Mm.}
  \label{fig:CS_different_res}
\end{figure*}

\section{The data-constrained MHD Equilibrium Model}
\label{sec:model}

{We seek MHD equilibrium consistent with a given snapshot of the
magnetic field observed in the photosphere (i.e., one single magnetogram)}. Such equilibrium is
assumed to exist when the corona is not in eruptive
stage. This is because the photospheric driving motions are so slow
that at any instant the coronal field has enough time to relax to a new
equilibrium. We solve the full set of MHD equations with the magnetic
field on the bottom boundary constrained by vector magnetogram from
SDO/HMI. Starting from a potential field model and an initially
hydrostatic plasma, we change the transverse field on the bottom
boundary incrementally until its matches the vector magnetogram. This
will drive the coronal magnetic field to evolve away from a potential
state. Once the bottom field is identical to the vector magnetogram,
the system is then let to relax to equilibrium with the bottom field
fixed.
The basic settings in this paper are similar
to our previous works~\citep{Jiang2012c, Jiang2013MHD,
  Jiang2016NC, Jiang2017}.
{Both viscosity and a small friction are used for the aid of the relaxation
process~\citep[see][]{Jiang2012c}}.  Here, the background plasma is initialized in a
hydrostatic, isothermal state with $T=10^{6}$~K (sound speed
$c_{S}=128$~km~s$^{-1}$) in solar gravity. Its density is configured
to make the plasma $\beta$ as small as $2\times 10^{-3}$ (the maximal
Alfv{\'e}n $v_{\rm A}$ is $4$~Mm~s$^{-1}$) to mimic the coronal
low-$\beta$ and highly tenuous conditions. The plasma thermodynamics
are simplified as an adiabatic energy equation since we focus on the
evolution of the coronal magnetic field. The bottom boundary of the
model is assumed as being the coronal base, thus the magnetic field
measured on the photosphere is used as a reasonable approximation of
the field at the coronal base. We used the solar surface magnetic
field data from the SDO/HMI~\citep{Schou2012HMI}, in particular, the
Space weather HMI Active Region Patches (SHARP) vector magnetogram
data series \citep{Hoeksema2014, Bobra2014}. Before input to the MHD
model, a Gaussian smoothing of the original data with FWHM of 2~arcsec (i.e., 1.4~Mm)
is used to remove the very small-scale features that cannot be
properly resolved by the MHD calculation.

The numerical scheme is an AMR-CESE-MHD code described
in details by~\citet{Jiang2010}. We use a computational
volume much larger than the region of interest for the
purpose of reducing the influences from {the side and top boundaries, where all the
variables are fixed as their initial values. At the bottom boundary, the velocity is fixed as zero.}
With an adaptive-mesh-refinement (AMR) technique, the computational
time is significantly saved while the resolutions for important structures
are still preserved.
In particular, the AMR is
designed to automatically capture narrow layer with strong currents as
well as resolve the strong magnetic field region (e.g., sunspot
regions) with strong gradients, which is illustrated in
\Fig~\ref{fig:amr_mesh}. Here we would like to emphasize that in
numerical sense, a CS is not a 2D surface but a narrow current layer
with thickness close to the grid resolution. For a typical resolution
of 1~Mm, the thickness of a CS should be less than a few Mm. For
capturing such intense current layer, the value of $J/B$ is used to
guide the refinement of the mesh. $J/B$ is proved to be a better
indicator that can highlight current sheet-like distribution than the
current density $J$ itself~\citep{GibsonFan2006, Fan2007,
  Jiang2016ApJ}. This is because in numerical realization, the CS
usually have both larger current density and weaker field than its
neighborhoods. As can be seen in \Fig~\ref{fig:amr_mesh},
the $J/B$ value in the CS abruptly increases over its neighborhoods, and thus
it is captured by the mesh points with highest resolution. During the
calculation, any location with $J/B$ becoming larger than $0.2/\Delta$
(where $\Delta$ is the local grid size) will be refined by a factor of
two. In addition, any place with strong magnetic field gradient or
strong current will be also refined, and the criteria are respectively
given by $|\nabla (B^{2}/2)|\Delta/\rho > 100 $ and
$|(\vec B\cdot\nabla)\vec B|\Delta/\rho > 100$ (where $\rho$ is the
plasma density, and all the variables are in normalized values). The
mesh shown in \Fig~\ref{fig:amr_mesh} used 6 levels of refinement with
highest resolution of 0.36~Mm in $x-y$ plane and 0.18~Mm in $z$ direction.


\section{Results}
\label{sec:results}


Three X-shaped homologous flares occurred in AR 11967 on 2014 February
2, when the region is close to the central meridian. As an example,
\Fig~\ref{fig:4aia_hmi} shows SDO observations of the third flare
at 18:10~UT. All these flares occurred in the same location
of the AR and demonstrated very similar morphology, i.e., X-shaped
ribbons and X-shaped brightening loops. As can be seen in
\Fig~\ref{fig:4aia_hmi}, the X extends long ``arms'' and has ``legs'' resembling those of
a running man. The flare ribbons {barely exhibit any movement
during the whole flaring phase}. All these flares are M-class and they
are confined without being associated with a coronal mass ejection or
a jet.

From the HMI magnetogram taken immediately prior to the flare, it is
suggested that a quadrupole magnetic configuration is responsible for
producing the X-shaped flares~\citep{LiuR2016NatSR, Kawabata2017}. As
shown in {\Fig~\ref{fig:4aia_hmi}c}, the two legs of the X-shaped
ribbon extended into two major sunspots with the inverse magnetic
polarities, labeled as P1, N1, while its arms extended into the
plage regions (labeled as P2, N2). Moreover, after the main phase of
the flares, hot loops of all four types of connections, i.e., P1-N1,
P2-N2, P1-N2 and P2-N1 were seen in AIA~131 and 94 channels,
suggesting the magnetic reconnection in the coronal magnetic
structures formed above the quadrupole magnetic polarity distribution.

In its lowest energy state, the coronal magnetic field can be modeled
by potential field. Based on a potential field extrapolation from the
vertical component of the vector magnetogram, we find that there is a
double-null magnetic topology of the quadrupole which confirmed the
finding of~\citet{LiuR2016NatSR}, who attempted to elaborate
the magnetic topology accounting for the same flare with {both potential-field
and NLFFF extrapolations}. In {\Fig~\ref{fig:4aia_hmi}e and f} we
show the locations of the null and the skeleton magnetic field lines
that delineate the magnetic separatrix. A well-defined X shape of
the field lines is seen, but obviously it does not match the observed
one, and both the nulls are situated far away from the center of the observed X-shaped
structure (with a distance of at least 30~Mm).
This indicates that the non-potentiality, that is, a stress of the
potential field and modification of the magnetic topology by the
electric currents in the corona, plays a key role in shaping the flare~\citep{Kawabata2017}.
{However, as shown in \citet{LiuR2016NatSR}, an NLFFF extrapolation produces a result even worse than the potential
field model in matching the flare topology}.

We use the vector magnetogram of the same time 18:00~UT as input to
the MHD model. Different from the potential model, the MHD model
identifies no coronal null but reveals a CS in the corona. \Fig~3 shows
the structure of the CS. In 3D, if defined by the iso-surface of
$J/B=1$~Mm$^{-1}$, the CS is a thin current layer with thickness of
$\sim 1.4$Mm. It is located in the center of the quadrupolar
configuration, and extends vertically all the way from the lower
boundary of the model box to a height of 40~Mm. The presence of the CS
is also prominent in the distribution of the current density $J$~(see
\Fig~\ref{fig:CS_structure}c), where the volumetric currents are
distributed much more smoothly in significantly larger space than the
CS. The CS separates field lines into two distinct connections, since when tracing field
lines from middle of the iso-surface, they clearly fall into two groups on either side of the layer
(see \Fig~\ref{fig:CS_structure}b), one connecting P1-N2, and the other connecting
P2-N1. The field lines naturally form an X-shaped
configuration. Moreover, by tracing all the field lines that are in
contact or pass through the CS, their footpoints on the bottom surface
delineate a sound X shape. This X shape, with its center, arms and legs,
almost coincides with the observed one in different AIA channels
(see \Fig~\ref{fig:ribbon_compare_qsl}). This strongly
suggests that reconnection triggered at the CS produces the flare,
since the locations of chromospheric flare
ribbons are well recognized to correspond to the footpoint locations of those magnetic field
lines that undergo reconnection in the CS~\citep{Qiu2009}. We further compute
a magnetic squashing-factor map~\citep{Titov2002} at the bottom
surface (see \Fig~\ref{fig:ribbon_compare_qsl}c and d), which is a useful tool to reveal all the
magnetic separatrix and quasi-separatrix layers (QSLs). The X-shaped
footpoints are co-spatial with two intersecting QSLs, which means
there is a HFT and the CS is formed at the HFT~\citep[see
also][]{LiuR2016NatSR}. We note that the squashing-factor map can
locate all the potential places for reconnection,
but it cannot tell
where the reconnection will actually take place for a flare. As can be
seen, the structure of QSLs is much more complex than that of the flare
ribbons~{\citep{Savcheva2015, Inoue2016}}. Thus, if ribbon locations are not known in advance, it is
still problematic to identify from all the QSLs the particular
flare-related one. Here, our model very directly shows the location
of the reconnection site for the flare.

Capturing of the CS is robust by our MHD model. For the other two
X-shaped flares, we can achieve similar results, i.e., the presence of CS and the
good match of the flare ribbons with footpoints of the field lines
contacting the CS~\citep[not shown here, but see
][]{YanX2017inprep}. Furthermore, the CS's location is not
sensitive to the resolution of grid, but its thickness and
intensity depend on the grid size. As shown in \Fig~\ref{fig:CS_different_res}, which gives
results for three different grid resolutions, the location of the
large value of $J/B$ (or $J$) is almost the same, while the peak value
increases approximately in proportion to the increasing of resolution.
Presumably, the current-density distribution
will approach to a $\delta$ function center at the CS, i.e.,
a magnetic tangential discontinuity, if the grid size
approaches infinitesimal, but before that the microscopic
behavior of plasma must be considered, which is beyond the scope of this paper.

\section{Conclusion}

In the paper we demonstrated the power of data-constrained MHD model
in reconstructing CS in the corona that is associated with flares. The
studied event is an atypical confined flare with X shape occurring in a
magnetically quadrupolar region. Neither the potential field model or
nonlinear force-free model can sufficiently reproduce the magnetic
topology in correspondence with the geometry of flare~\citep[see][]{LiuR2016NatSR}.
Our results show actually there is a large-scale CS
formed prior to the flares, and the field lines traced from the CS to
the photosphere form an X shape that rather precisely reproduces the
geometry of the flare ribbons. We thus suggest that the observed
X-shaped flare is a product of the dissipation of the CS through
reconnection. The recurrence of the X-shaped flares {might} be attributed
to the repetitive formation and dissipation of the CS, {while the
formation of CS is driven by the photospheric footpoint motions~\citep{Santos2011}}. Such
a dynamic process will be further investigated by a data-driven MHD
model~\citep[e.g.,][]{Jiang2016NC, Jiang2016ApJ}.

\acknowledgments

This work is supported by the
National Natural Science Foundation of China (41574170, 41574171, 41531073,
41374176,  41231068, and 11373066), Shenzhen Technology Project JCYJ20170307150645407,
and the Specialized Research Fund for State Key Laboratories.
Data from observations are courtesy of NASA {SDO}/AIA and the HMI science teams.
We thank Rui Liu and Jun Chen for providing us the code for calculating the magnetic
squashing factor.


\end{CJK*}
\end{document}